\documentclass[ preprint, twocolumn, number, 5p]{elsarticle}
\makeatletter
\def\ps@pprintTitle{%
 \let\@oddhead\@empty
 \let\@evenhead\@empty
 \def\@oddfoot{}%
 \let\@evenfoot\@oddfoot}
\makeatother
\usepackage{lineno}
\usepackage{hyperref}
\usepackage{amsmath}
\usepackage{caption}
\usepackage{subcaption}
\usepackage{xcolor}
\usepackage{float}
\usepackage{times}
\usepackage{pgfplots}  
\usepackage{tikz}  

\graphicspath{{figs/}} 

\pgfplotsset{compat=1.14}
\usepgfplotslibrary{groupplots, fillbetween}
\usetikzlibrary{calc, angles, positioning, intersections, decorations.pathreplacing, quotes}

\newcommand{\sbar}{\overline{s}}
\newcommand{\FigBeg}{Figure~}
\newcommand{\FigMid}{Fig.~}

\newcommand{\EqMid}{Eq.~}
\newcommand{\phii}{\varphi_{\textrm{ind}}}
\newcommand{\phig}{\varphi_{\textrm{g}}}
\newcommand{\Ng}{N_{\textrm{g}}}
\newcommand{\sstar}{s^{*}}
\newcommand{\spert}{s_{\textrm{pert}}}

\newcommand{\sgn}{\mathrm{sgn}}

\begin{document}

\begin{frontmatter}

\title{A Dynamical Model for the Origin of Anisogamy}

\author[ESAMaddress]{Joseph D. Johnson\corref{mycorrespondingauthor}}
\ead{JosephJohnson2020@u.northwestern.edu}
\author[ESAMaddress]{Nathan L. White}
\author[ESAMaddress]{Alain Kangabire}
\author[ESAMaddress,Physaddress,NICOaddress]{Daniel M. Abrams}

\address[ESAMaddress]{
Department of Engineering Sciences and Applied Mathematics, Northwestern University, Evanston, IL, 60208, USA}
\address[Physaddress]{Department of Physics and Astronomy, Northwestern University, Evanston, IL, 60208, USA}
\address[NICOaddress]{Northwestern Institute on Complex Systems, Evanston, IL, 60208, USA}
\cortext[mycorrespondingauthor]{Corresponding author}

\begin{abstract}
The vast majority of multi-cellular organisms are anisogamous, meaning that male and female sex cells differ in size. It remains an open question how this asymmetric state evolved, presumably from the symmetric isogamous state where all gametes are roughly the same size (drawn from the same distribution). Here, we use tools from the study of nonlinear dynamical systems to develop a simple mathematical model for this phenomenon.  Using theoretical analysis and numerical simulation, we demonstrate that competition between individuals that is linked to the mean gamete size will almost inevitably result in a stable anisogamous equilibrium, and thus isogamy may naturally lead to anisogamy.
\end{abstract}


\end{frontmatter}


\section{Introduction}

``Anisogamy'' refers to the observation that gamete size distributions in many species are bimodal or multimodal, and has long been a topic of study (see, e.g., \cite{kalmus1932erhaltungswert,kalmus1960evolutionary,scudo1967adaptive,parker1972origin,bell1978evolution,cox1985gamete,hurst1990parasite,bonsall2006evolution,blute2013evolution,lehtonen2016anisogamy}). Anisogamy is common in complex organisms such as plants, animals, fungi, and certain algae (\cite{parker1972origin,haig1988model,billiard2011having,bateman1994heterospory}). There is a consensus in the literature  that anisogamy evolved from isogamy, where sexual reproduction occurs between sex cells that are the same size (\cite{parker1972origin,bell1978evolution,bulmer2002evolution,hayward2011cost}).

Anisogamy has been theorized to be a factor in the development of differences between sexes. Bateman credits to anisogamy the fact that male \textit{Drosophila melanogaster} are far more eager than females to mate  (\cite{bateman1948intra}). Lehtonen et al.~add theory to this intuition, demonstrating that, as the size ratio between large and small gametes increases, organisms with small gametes will choose to allocate more resources to searching for mates and warding off others with small gametes from potential mates (\cite{lehtonen2016anisogamy}).

A related question that remains of scientific interest is why most complex organisms have only two sexes.   This is the case for almost all animals, but, e.g., fungi may have scores or even thousands of ``mating types'' (the term ``sex'' is typically not used in this case) (\cite{kues1993origin,billiard2011having,kuees2015two}). We do not directly address this here, but a better understanding of the origin of anisogamy might also inform our understanding of this question. 

Various attempts have been made to explain the evolution of anisogamy. \cite{kalmus1932erhaltungswert} and \cite{scudo1967adaptive} argue that the number of successful fusions is maximized when the difference between gamete sizes is vast.

\cite{parker1972origin} posit that anisogamy developed due to disruptive selection acting on an isogamous population where gamete size has an inverse relationship with gamete production and a positive relationship with zygote viability. The authors argue that this transition from isogamy to anisogamy depends on how zygote fitness varies with zygote volume.  \cite{bell1978evolution} and \cite{charlesworth1978population} follow up the work done by Parker et al.~by giving an analytical framework for this theory, further illuminating the relationship between the scaling of fitness with respect to size and the development of anisogamy. 

\cite{bulmer2002evolution} expand on this approach by factoring in explicit survival function for gametes and zygotes. They demonstrate that shifting the zygote survival function while keeping the gamete survival function fixed  can lead to the development of anisogamy. The authors also adjust their model for the existence and stability of an anisogamous evolutionary stable strategy (ESS) given a critical minimal gamete size. 

Our approach differs from most prior work in three key ways. (1) We do not assume fusions occur only between dissimilar gametes (i.e., no mating types), (2) we assume that the viability of a gamete is not determined by absolute size, but rather by the difference from the gamete population mean (a form of frequency-dependence), and (3) we move outside the framework of the ESS and use dynamical systems theory to find the conditions for anisogamy.

\section{Model development}

In this section we begin with a concrete version of our model using specific algebraic functions; in Section \ref{sec:geo_arg} we generalize our analysis to arbitrary functions with some known limiting properties. For simplicity we develop our model under the assumption that gametes fuse randomly; this assumption could likely be relaxed (though we leave that for future work) but will remaining an operating assumption here. We further assume that mating types do not exist, and hence that there are no restrictions on which gametes can mate with which. \footnote{We believe that the evolution of mating types need not occur simultaneously with the emergence of anisogamy, but leave explicit study of that question for future work.}

\subsection{Individual reproductive potential}

Consider a population of $N$ organisms with gametes that have sizes $s_j$, $j=1, \ldots, N$. Following the approach used in \cite{clifton2016}, we denote the individual ``reproductive potential'' of the $j$th individual by $\phii$, defined as some increasing function of the fitness (the expected number of adult offspring it will produce)\footnote{For a brief discussion of the relationship between fitness and reproductive potential, see \ref{sec:potfit}.}.  We assume that this potential can be expressed as a product of $\Ng$, the expected number of gametes produced, and $\phig$, the average reproductive potential of its gametes (where gamete reproductive potential is, similarly, an increasing function of gamete fitness---the expected number of adults resulting from that gamete, with upper bound 1, ignoring monozygotic twinning):
\begin{equation}
\phii = \Ng \phig \;. \label{eq:ind_fit_gen}
\end{equation}
Because we are concerned with anisogamy and hence gamete size distributions, we ignore all factors influencing reproductive potential besides gamete size.  Other factors are clearly extremely important, but we model only the effects of gamete sizes on  reproductive potential here, and thus write that $\Ng = \Ng(s_j)$, $\phig = \phig(s_j)$.

\subsection{Gamete production function} \label{sec:gamete_prod}

We assume that $\Ng$ is a decreasing function of gamete size due to the fact that each organism has limited resources (physical, temporal and energetic) to dedicate to gamete production. Some observational evidence supports this: smaller male sex cells are far more numerous than significantly larger female sex cells (\cite{wallace2010human, johnson1983further, alberts2002eggs, bellastella2010dimensions}); additionally, research has found a negative relationship between clutch size and egg size in the black-backed gull \textit{Larus fuscus} (\cite{nager2000within}), and across various species of snakes (\cite{hedges2008lower}). 

To present a concrete analytical argument, and motivated by their ubiquity in nature (\cite{brown2002fractal, newman2005power, clauset2009power}), we choose $\Ng$ to be a power law, i.e.,
\begin{equation}
\Ng(s_j) = c_1 s_j^{-\alpha}
\label{eq:num_gam}
\end{equation}
where $c_1$ is a constant of proportionality and the constant $\alpha$ is assumed to be positive. In the section ``\nameref{sec:geo_arg}'' below, we generalize our argument to arbitrary decreasing functions.

\subsection{Gamete reproductive potential}  \label{sec:gamete_fit}
We assume that $\phig$ is an increasing function of gamete size.  This is motivated by the idea that increased size indicates increased provisions to promote survival of the gamete and the zygote potentially formed after fusion with another gamete. Some evidence supports this link: associations between between egg size (measured by volume or mass) and positive offspring outcomes have been reported in various avian species (\cite{blomqvist1997parental, krist2011egg, erikstad1998significance, valkama2002inter}).

Critically for our model, we assume that the fitness ``payoff'' accruing to larger gametes is relative rather than absolute in nature.  That is, we assume that a gamete of size $s_j$ will have greater reproductive potential in a population where it is among the largest than in a population where it is among the smallest.  This assumption (a form of frequency-dependent selection) is motivated by the hypothesis of zygote competition, and ultimately by the same idea underlying natural selection: if environmental conditions preclude all viable zygotes from reaching adulthood, those with the greater provisions afforded by larger parental gamete sizes will be more likely to survive.  A similar argument can be made if direct competition between gametes plays a role in determining fitness.

Thus, we link the reproductive potential of the $j$th gamete to the full distribution of gamete sizes in the population.  We can express such a link in simple terms by assuming $\phig(s_j)$ is an increasing function of $s_j-\sbar$, where $\sbar = N^{-1} \Sigma_{j=1}^N s_j$ is the mean gamete size in the population.

We expect reproductive potential to saturate for both extremely large and extremely small gametes, so we choose a sigmoidal form for our analytical expression of $\phig(s_j)$:
\begin{equation}
\phig(s_j \vert \sbar) = c_2 \left( 1 + \frac{s_j - \sbar}{w + \left| s_j - \sbar \right|} \right)\;, 
\label{eq:gam_fit}
\end{equation}
where $c_2$ is a constant of proportionality and $w$ sets the width of the sigmoid. In the section ``\nameref{sec:geo_arg}'' below, we generalize our argument to a wider class of functions $\phig(s_j-\sbar)$. 

Substituting \EqMid \eqref{eq:num_gam} and \EqMid \eqref{eq:gam_fit} into \EqMid \eqref{eq:ind_fit_gen}, we obtain the following individual reproductive potential function: 
\begin{equation}
\phii(s_j \vert \sbar) = \Ng(s_j)  \phig(s_j \vert \sbar) = c_3 s_j^{-\alpha} \left( 1 + \frac{s_j - \sbar}{w + \left| s_j- \sbar \right| } \right)\;,
\label{eq:ind_fit}
\end{equation}
where we have combined the multiplicative constants of proportionality into a single constant $c_3 = c_1 c_2$.

\subsection{Gamete size evolution}

We assume that natural selection acts on the population in such a way that gamete sizes change at a rate proportional to the reproductive potential to be gained. That is, there is a ``phenotype flux'' 
\begin{equation}
\frac{ds_j}{dt} =  \frac{1}{\tau} \frac{\partial \phii}{\partial s_j},\quad j=1,\ldots,N\;.
\label{eq:diff_ind_fit}
\end{equation}
In the continuum limit $N \to \infty$, these $N$ ordinary differential equations are replaced by a single partial differential equation---the continuity equation for $\rho(s,t)$, the probability density function describing the distribution of gamete sizes:
\begin{equation}
\dfrac{\partial \rho}{\partial t} =  -\nabla \cdot \left( \dfrac{ds}{dt}\rho \right)\;, 
\label{eq:cont_eq}
\end{equation}
where $ds/dt$ is given by
\begin{equation}
\frac{ds}{dt} =  \frac{1}{\tau} \frac{\partial \phii}{\partial s}\;.
\label{eq:diff_ind_fit2}
\end{equation}
Here $\tau$ sets the time scale for the evolution of gamete size.  Since this is unknown (and not the focus of this work), we rescale time such that $\tau=1$ without loss of generality. 

To be clear, we are not assuming that individual organisms explicitly change their gamete sizes in this model, rather, the ``phenotype flux'' $ds/dt$ captures how the gamete size distribution $\rho(s,t)$ changes over long time scales. Note that probability density functions such as $\rho(s,t)$ must obey the continuity equation (Eq.~\eqref{eq:cont_eq}).
In \cite{clifton2016}, the authors demonstrated how this approach (substitution of the phenotype flux from \EqMid \eqref{eq:diff_ind_fit2} into the continuity equation) can be considered equivalent to a ``replicator equation'' approach (\cite{taylor1978evolutionary,schuster1983replicator}) for appropriate choices of fitness functions.  We summarize this connection between the continuity connection and the replicator equation in \ref{sec:potfit}.

\section{Model implications} \label{sec:results}

\subsection{Existence of the anisogamous equilibrium} 
\begin{figure}[t!]
	\centering
	\begin{tikzpicture}
	\begin{axis}[width = 88mm, height = 62mm, 
	xmin=0, xmax=3.5,
	ymin=0, ymax=1.75,
	axis lines=center,
	clip =false,
	axis line style = ultra thick,
	axis on top=true,
	domain=-3.5:3.5,
	xtick = {0.001,1,2,3},
	xticklabels={0,,$\sbar$,},
	ytick = {0.001},
	yticklabels={},
	tickwidth=3mm
	]
	\addplot [name path=F3, samples=2, black, ultra thick, domain = 0:0.0277]{0};
	\addplot [name path=ka3, samples=2, black, ultra thick, opacity=0, domain = 0:0.0277]{1.75};
	\addplot [name path=F, samples=2, black, ultra thick, domain = 0.0277:1]{0};
	\addplot [name path=ka, samples=50, black, ultra thick, domain = 0.0277:1]{ \x^(-1)*(1+10*(\x-2)*(1+10*abs(\x-2))^(-1)) };
	\addplot [name path=F1, samples=2, black, ultra thick, domain = 1:2.26]{0};
	\addplot [name path=ka1, samples=50, black, ultra thick, domain = 1:2.26]{ \x^(-1)*(1+10*(\x-2)*(1+10*abs(\x-2))^(-1)) };
	\addplot [name path=F2, samples=2, black, ultra thick, domain = 2.26:3.5]{0};
	\addplot [name path=ka2, samples=50, black, ultra thick, domain = 2.26:3.5]{ \x^(-1)*(1+10*(\x-2)*(1+10*abs(\x-2))^(-1)) };
	\addplot+[red!30] fill between[ of = F and ka];
	\addplot+[blue!30] fill between[ of = F1 and ka1];
	\addplot+[red!30] fill between[ of = F2 and ka2];
	\addplot+[red!30] fill between[ of = F3 and ka3];
	\addplot+[mark=none, black, ultra thick, dotted] coordinates {(2, 0) (2, 1.5)};
	\draw [black, fill] (axis cs:2,1.5) circle (0pt) node [above] {$s_j = \sbar$};
	\draw[] (3.65,0) node{$s_j$};
	\draw[] (-0.2, 0.85) node{\rotatebox{90}{Ind. reproductive potential (a.u.)}};
	\draw[] (1.75,-0.32) node{Gamete size $s_j$};
	\draw[] (2.275,-0.153) node{$\sstar$};
	\draw[] (0, 2) node{$\varphi(s_j\vert\sbar)$};
	\end{axis}
	\end{tikzpicture}
	\caption{{\bf Example individual reproductive potential function.}
		Here we show the reproductive potential function defined by \EqMid \eqref{eq:ind_fit} in arbitrary units (a.u.). Two maxima are apparent, one at zero and another at a nonzero value $\sstar$. Dynamics given by \EqMid \eqref{eq:diff_ind_fit} are illustrated by color with red indicating regions where gamete size decreases and blue indicating where gamete size increases.  For this illustration, we set $w = 1/10$, $\alpha = 1$, and $\sbar = 2$.  }
	\label{fig:ind_fit}
\end{figure}
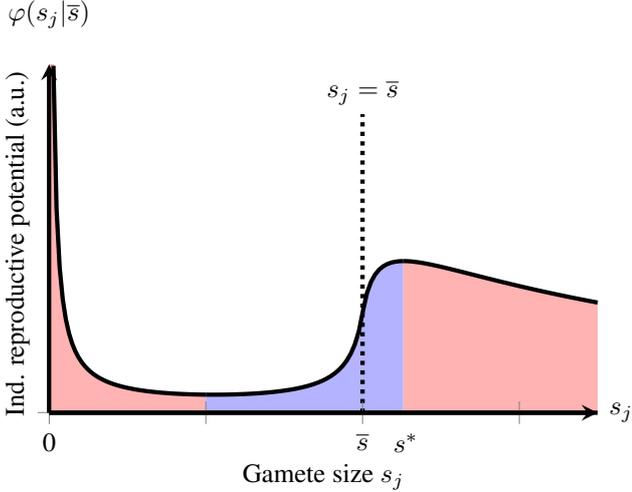

The development of anisogamy through intraspecific competition can be seen as a form of disruptive selection (\cite{da2018evolution}). This implies a fitness landscape with multiple distinct peaks.   

For $\phii$ as defined in \EqMid \eqref{eq:ind_fit}, at most two local maxima can exist: one at $s_j = 0$ and another at a nonzero value $\sstar$. This is illustrated in \FigMid \ref{fig:ind_fit}, where two local maxima can be seen.
We therefore assume that the anisogamous equilibrium takes the form $\rho(s) = x \delta(s - 0) + (1-x) \delta(s-\sstar)$,\footnote{See \ref{sec:nonzero_gamete} for the case when the small gamete group has finite, nonzero equilibrium gamete size.} where $\delta$ is the Dirac delta function and $0 < x < 1$ is the proportion of gametes that are small (i.e., the proportion that might be referred to as primitive ``male'' gametes). This equilibrium must be self-consistent, meaning that the first moment of the distribution is indeed the same as the average gamete size $\sbar$. Substituting $\rho(s,t) = x \delta(s - 0) + (1-x) \delta(s-\sstar)$, $\sbar = (1-x) \sstar$, \EqMid \eqref{eq:ind_fit}, and \EqMid \eqref{eq:diff_ind_fit2} into \EqMid \eqref{eq:cont_eq} and solving $\partial \rho / \partial t = 0$ (or, equivalently, setting $\partial \phii/\partial s = 0$ after plugging \EqMid \eqref{eq:ind_fit} into \EqMid \eqref{eq:diff_ind_fit2}, with $\sbar = (1-x) \sstar$), we find
\begin{equation}
\sstar =  w \frac{1 - 3 \alpha x + \sqrt{\beta}}{4 \alpha x^2}\;, 
\label{eq:egg_size}
\end{equation}
where $\beta = \alpha^2 x^2 - 6 \alpha x + 1$. 
An anisogamous equilibrium thus exists for all positive $\alpha$ and $w$, as long as $\beta > 0$.\footnote{See \ref{sec:sexratios} for discussion of the implications of this requirement.}

\subsection{Stability of the anisogamous equilibrium} \label{sec:stab}

Consider the perturbation of a single individual from the large gamete group by an amount $\epsilon \ll 1$ in the limit $N \gg 1$, so this represents an infinitesimal change to the full gamete distribution.  We set $\sbar = (1-x) \sstar$ and $\spert = \sstar + \epsilon$, where $\sstar$ is given by \EqMid \eqref{eq:egg_size}.
Substituting into \EqMid \eqref{eq:diff_ind_fit2} and Taylor expanding to linear order in $\epsilon$, we find 
\begin{equation}
\frac{d\epsilon}{d t} = Q(x,\alpha) \epsilon \;,
\label{eq:large_stab}
\end{equation}
where
\begin{align}
Q = -\frac{16 x \alpha^2 (\sstar)^{-1-\alpha} (1-x+\sqrt{\beta})}{w (1 + x \alpha +\sqrt{\beta})^2}  \;, 
\end{align}
with $\beta$ defined as in \EqMid \eqref{eq:egg_size}.  Here $Q < 0$ for all allowable parameter values, and thus the anisogamous state is stable under this kind of perturbation. 

A similar perturbation of one individual from the small gamete group is simply $\spert = \epsilon$, which, when substituted into \EqMid \eqref{eq:diff_ind_fit2} yields
\begin{equation}
\frac{d\epsilon}{d t} = -c_3 \alpha \frac{w}{w+\sbar} \epsilon^{-1-\alpha} 
\label{eq:small_stab}
\end{equation}
when truncated at leading order.  Since $c_3$, $\alpha$, $w$, and $\sbar$ are all positive, the anisogamous state is stable to infinitesimal perturbations of this sort whenever it exists.  

We omit a more general examination of stability, but in  \ref{sec:linear_stab_gam} we show that all eigenvalues of the finite $N$ system are negative for $N \gg 1$, and thus that the anisogamous state is indeed linearly stable.

\subsection{Geometric argument} \label{sec:geo_arg}

For clarity and convenience, we earlier assumed specific algebraic forms for $\phig$ and $\Ng$. We now show the possible emergence of anisogamy in a system where only the asymptotic properties of those functions are known.

We start by expanding the derivative on the right-hand side of \EqMid \eqref{eq:diff_ind_fit2}:
\begin{equation}
\frac{ds}{dt} =  \frac{1}{\tau} \dfrac{\partial \phii}{\partial s} = \frac{1}{\tau} \left[  \frac{\partial \phig}{\partial s} \Ng + \frac{d \Ng}{ds} \phig \right] \;.
\end{equation}
At an equilibrium $s=\sstar$, the net phenotype flux $ds/dt = 0$.  Assuming  $\Ng>0$ and $\phig>0$, we find that the following condition must hold at each $\sstar$:
\begin{equation}
\dfrac{\phig'}{\phig}  = -\dfrac{\Ng'}{\Ng}
\label{eq:geo_arg}
\end{equation}
where $' \equiv \partial/\partial s$. 

The left side of \EqMid \eqref{eq:geo_arg} is the relative change in gamete reproductive potential and the right side is the magnitude of the relative change in gamete production. Gamete sizes will increase when the reproductive potential gains outweigh the decreased gamete production, and will shrink when the opposite is true.

\begin{figure}[t!]
	\centering
	\begin{tikzpicture}
	\begin{axis}[width = 88mm, height = 62mm, 
	xmin=0, xmax=3.5,
	ymin=0, ymax=3.5,
	axis lines=center,
	clip =false,
	axis line style = ultra thick,
	axis on top=true,
	domain=-3.5:3.5,
	yticklabels={,,},
	xticklabels={,,},
	tickwidth=0mm
	]
	\addplot [name path=F, samples=100, red, ultra thick, domain =0:3.45]{2/(cosh(\x))^2};
	\addplot [name path=ka, samples=100, blue, ultra thick, dotted, domain = 0:3.45]{ 3*(\x+1)^(-2.5)};
	\draw[] (1.75,-0.25) node{Gamete size $s$};
	\draw[red] (1,1.5) node{$\phig'/\phig$};
	\draw[blue] (0.35,0.55) node{$-\Ng'/\Ng$};
	\draw[] (3.7,0) node{$s$};
	\end{axis}
	\end{tikzpicture}
	\caption{{\bf Geometric argument for anisogamy.}
		We illustrate a case where gamete reproductive potential, $\phig$, and gamete production, $\Ng$, satisfy the conditions set out in the section ``\nameref{sec:geo_arg}.'' When gametes are small, the relative gains due to the ability to produce more of them $|\Ng'/\Ng|$ outweigh the relative drop in reproductive potential $|\phig'/\phig|$.  In some intermediate range, reproductive potential gains dominate, and then as gametes become very large the production terms again dominate as reproductive potential gains saturate.}
	\label{fig:graph_arg}
\end{figure}
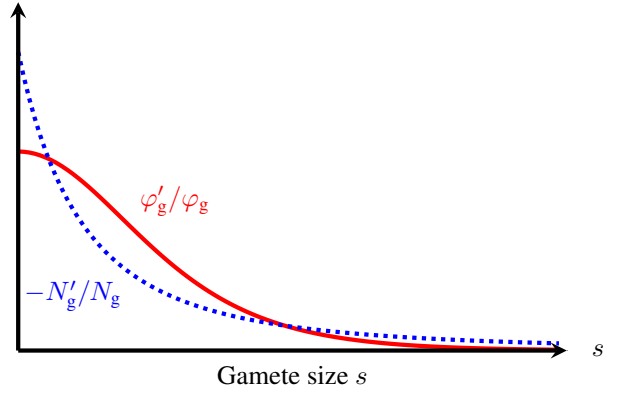

The existence of anisogamy requires that two distinct intersections must exist between the functions on the left and right-hand sides of \EqMid \eqref{eq:geo_arg} (see Fig.~\ref{fig:graph_arg}).  The following conditions are thus sufficient for anisogamy to exist: 
\begin{enumerate}
	\item Continuity of $\phig'$ and $\Ng'$.
	
	\item The gamete production terms dominate as size approaches zero (relative decrease in production larger than relative increase in reproductive potential), i.e.,
	\[ 
	\left| \frac{\phig'}{\phig}\right | <  \left| \frac{\Ng'}{\Ng} \right|, s \to 0^+ \;.
	\]
	This is reasonable if the potential saturates at some minimum  (possibly zero) for small gametes.
	\label{item:graph_condsa}
	
	\item There exists at least one finite value of $s$ (say $s=a$, $a>0$) at which reproductive potential terms dominate over gamete production terms, i.e., 
	\[ 
	\left| \frac{\phig'}{\phig}\right | >  \left| \frac{\Ng'}{\Ng} \right|, s = a \;.
	\]
	If this fails, smaller gametes are always better for fitness.  As long as there is some ``provisioning'' advantage to larger gametes at some point, however, this condition should be satisfied. 
	
	\item Gamete production terms again dominate as size goes to infinity, i.e., 
	\[ 
	\left| \frac{\phig'}{\phig}\right | <  \left| \frac{\Ng'}{\Ng} \right|, s \to \infty \;.
	\]
	This is reasonable if fitness gains eventually saturate.
\end{enumerate} 
In addition to the above, a self-consistency condition must also hold: It must be possible for the function $\phig' / \phig$ to satisfy 
\[ 
\left| \frac{\phig'}{\phig}\right | = \left| \frac{\Ng'}{\Ng} \right|\;,
\]
given $\sbar = (1-x)\sstar$, for some fractionation $x \in (0,1)$. 

\FigBeg \ref{fig:graph_arg} shows an example of functional shapes for $\phig$ and $\Ng$ that satisfy these conditions. Note that a wide variety of gamete reproductive potentials $\phig$ can work---the function need not be sigmoidal nor even monotonic.

\section{Results and discussion}

We have put forth a model that provides a plausible explanation for the development and stability of anisogamy, even without the existence of mating types.  This model is based upon the assumption that an individual's overall reproductive potential can be broken down into a ``gamete production'' term quantifying the number of gametes produced, and a ``gamete potential'' term quantifying those gametes' likelihood of eventually forming zygotes that reach adulthood.  Both of these are assumed to depend upon gamete size, with gamete reproductive potential having a positive relationship with gamete size, and number of gametes having a negative relationship with gamete size. A critical assumption is that size-dependence for gamete reproductive potential is determined relative to the mean of the population, encapsulating the intra-species competition for resources.

Although other models have been proposed to explain anisogamy, ours requires minimal assumptions and accounts for its emergence from an initially isogamous state.  We require no assumptions about the existence of mating types; we hope that competing theories with and without mating types may in the future be distinguished based on data.  For simplicity and clarity, we have treated individuals in this work as identical without variation, and we have allowed small gametes to approach zero size.  More realistic assumptions do not appear to change the broad results shown here (see Sections \ref{sec:Num_Sim_Gam}--\ref{sec:abs_gam_fit} for various numerical experiments).

\section{Acknowledgments}
The authors gratefully acknowledge support of the National Science Foundation through the program on Research Training Groups in the Mathematical Sciences, grant 1547394.  We also thank Northwestern University's Office of Undergraduate Research for support through URP 758SUMMER1915627 and URP 758SUMMER1915476.  We also thank Christina Goss for help with translation of reference \citep{kalmus1932erhaltungswert}.

\appendix

\section{Sex ratios} \label{sec:sexratios}
\begin{figure}[ht!]
	\centering
	\begin{tikzpicture}
	\begin{axis}[width = 88mm, height = 62mm, 
	xmin=0, xmax=4.0,
	ymin=0, ymax=1.00,
	axis lines = center,
	clip = false,
	axis lines = middle,
	axis line style = ultra thick,
	axis on top = true,
	domain = 0:4.5,
	xtick = {0.001,1,2,3,4},
	xticklabels={0,1,2,3,4},
	ytick = {0.001,0.25,0.5,0.75,1},
	yticklabels={0,,,,1},
	tickwidth=3mm
	]
	\addplot [name path=F, samples=200, black, ultra thick, domain = 0.172:4]{(3-2*2^(0.5))*(1/\x)};
	\addplot [name path=ka, samples=2, black, ultra thick, domain = 0.172:4]{ (0 };
	\addplot [name path=ka2, samples=2, black, dotted, ultra thick, domain = 0.172:4]{ 1 };
	\addplot [name path=ka3, samples=2, black, dotted, ultra thick, domain = 0:0.172]{ 1 };
	\addplot [name path=ka4, samples=2, black, ultra thick, domain = 0:0.172]{ (0 };
	\addplot+[red!30] fill between[ 
	of = F and ka2]; 
	\addplot+[blue!30] fill between[ 
	of = ka4 and ka3]; 
	\addplot+[blue!30] fill between[ 
	of = ka and F];
	\draw[] (0,1.05) node{$x$};
	\draw[] (2,-0.2) node{Power law exponent $\alpha$, where $(\textrm{\# gametes}) \propto \textrm{(size)}^{-\alpha}$};
	\draw[] (-0.4, 0.5) node{\rotatebox{90}{Fraction ``male'' $x$}};
	\draw[] (4.15,0) node{$\alpha$};
	\end{axis}
	\end{tikzpicture}
	\caption{\textbf{Possible sex ratios.} The solid black curve shows the threshold for existence of the anisogamous state given by \EqMid $\eqref{eq:bimod_cond}$. The anisogamous equilibrium exists below the threshold (blue shaded region) and ceases to exist above the threshold (red shaded region). Here the fraction ``male'' refers to the fraction with small gametes.}
	\label{fig:sex_ratio}
\end{figure}
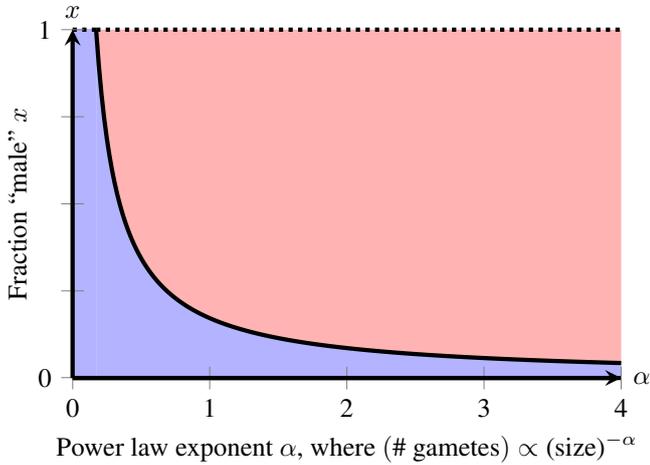


For the bimodal gamete size distribution to exist, $\sstar$ must be real-valued, and hence $\beta = \alpha^2 x^2 - 6 \alpha x + 1$ in \EqMid \eqref{eq:egg_size} must be positive.  This holds when
\begin{equation}
x < \frac{3-2\sqrt{2}}{\alpha} \;. \label{eq:bimod_cond}
\end{equation}
Thus there is an implied range of stable sex ratios for a given value of $\alpha$. \FigBeg \ref{fig:sex_ratio} illustrates the relationship between the power law exponent $\alpha$ and the fraction of the population with small gametes (the fraction ``male'') given by \EqMid \eqref{eq:bimod_cond}. As $\alpha$ increases in magnitude the range of possible fractionations decreases. 

Interestingly, an approximate 1:1 sex ratio is not attainable for some ``reasonable'' exponents of $N_g$ (e.g., 1, 2, and 3, each of which would correspond to a distinct simple measure of gamete ``size'').  This is likely a result of our specific choices of $\phig$ and $N_g$, as well as the restricted nature of the model.  When we modify gamete reproductive potential to depend on both relative and \textit{absolute} gamete size, in numerical simulation we observe stable anisogamous states with arbitrary sex ratios. Also note that our model purposefully omits frequency-dependent selection effects that would likely drive sex ratios toward 1:1 (the reproductive potential of a single ``male'' gamete in a community of mostly ``female'' gametes would be much higher than in a community of mostly ``male'' gametes because the likelihood of fusion would be higher and the likelihood of zygote survival would be higher, i.e., Fisher's principle \citep{hamilton1967extraordinary}).

\begin{figure}[th!]
	\centering
	\begin{tikzpicture}
	\node at (0,0) {\includegraphics[width =0.95\columnwidth]{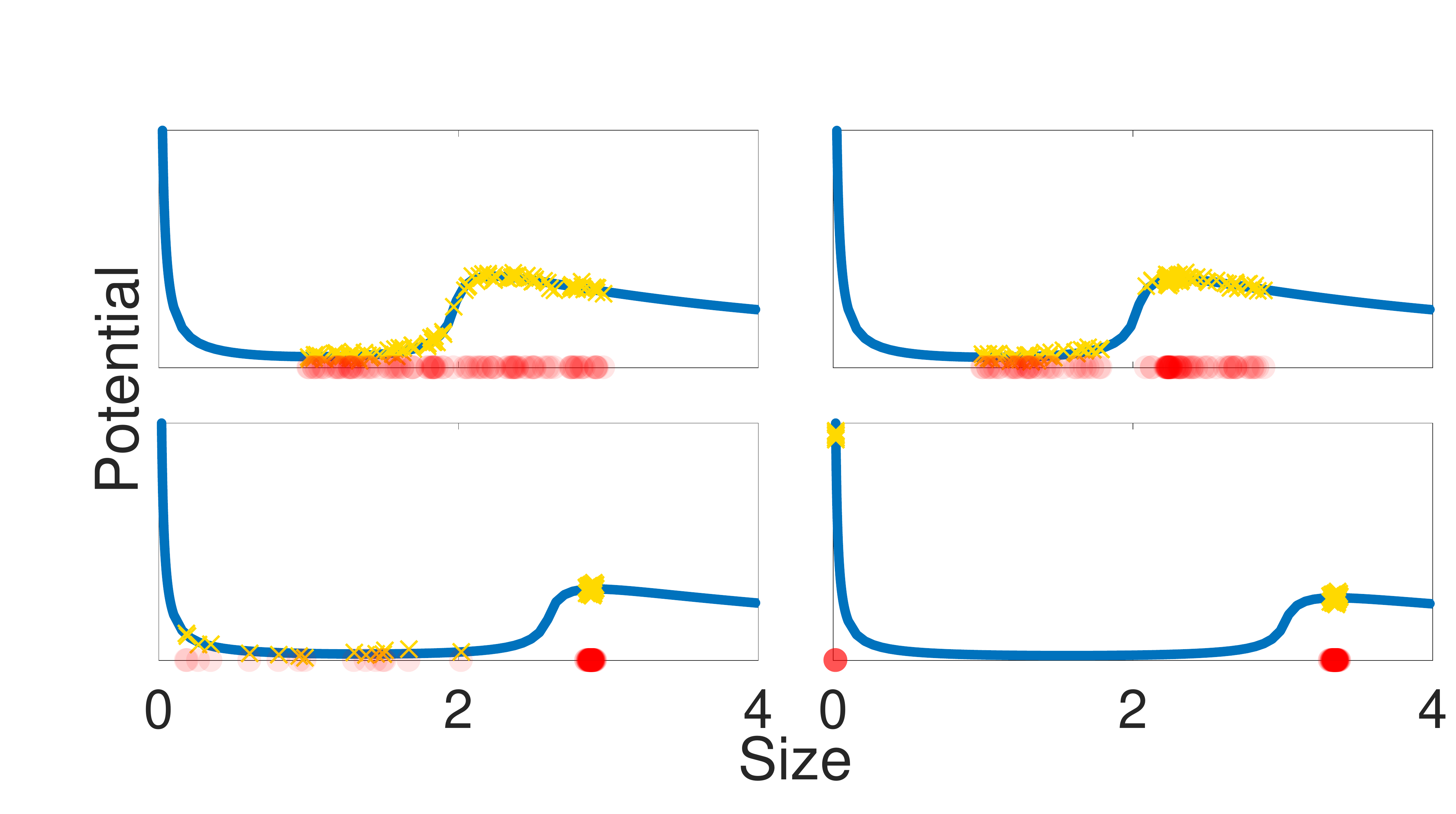}};
	\node at (-0.34,1.85) {(a)};
	\node at (3.85,1.85) {(b)};
	\node at (-0.34,0.05) {(c)};
	\node at (3.85,0.05) {(d)};
	\end{tikzpicture}
	\caption{\textbf{Simulation of the system.} Panels (a)-(d) show the evolution of the system from an isogamous state to an anisogamous state. Here, the blue curve shows the reproductive potential landscape given by \EqMid \eqref{eq:ind_fit}, the red circles indicate gamete sizes, and the yellow crosses give positions of gametes along the reproductive potential landscape. Panel (a) captures the isogamous initial condition $\mathcal{U}(1,3)$. Panel (b) shows the individuals moving along the landscape in the direction that increases reproductive potential. Panel (c) shows the beginning of two groups forming. In Panel (d), the simulation has arrived at an anisogamous equilibrium, with gamete sizes converging to zero or $\sstar$ as given by \EqMid \eqref{eq:egg_size}. The final fraction of organisms that produce small gametes is $x= 0.1$. For this numerical experiment, we set $\alpha = 1$, $N = 100$, and $w = 1/10$. }
	\label{fig:gamete_evo}
\end{figure}
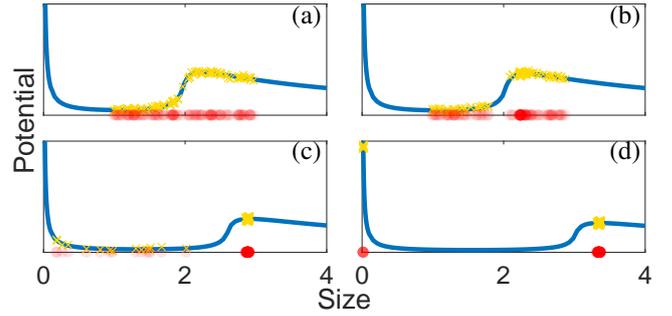

\section{Numerical simulations} \label{sec:Num_Sim_Gam}

We test predictions of our model via numerical simulation. \FigBeg \ref{fig:gamete_evo} shows the evolution of a population from a state that is isogamous to a one that is anisogamous. The gametes (yellow) move along the landscape (blue) in the direction that increases their reproductive potential. For this simulation, we set $\alpha = 1$, $N = 100$, and $w = 1/10$ with a final fraction of individuals producing small gametes $x = 0.1$.  The initial isogamous distribution was sampled from the uniform distribution $\mathcal{U}(1,3)$.

\section{Stability tests}

\subsection{Linear stability} \label{sec:linear_stab_gam}

Section \ref{sec:stab}, we outlined a restricted stability test of the anisogamous equilibrium where only single-gamete perturbations were allowed.  A more rigorous test of stability is difficult because the Dirac delta functions that comprise the equilibrium gamete size distribution are actually generalized functions and thus must be treated carefully when perturbed.  One straightforward way to avoid this difficulty is to look at the linear stability of the equilibrium for finite $N$, then take the limit as $N \to \infty$. 

One can show that, for finite $N$, the off-diagonal elements of the Jacobian matrix take the form
\begin{align}
J_{ij} =  (s_i)^{-\alpha-1}&\left[ 2\frac{N-1}{N^2} \frac{ w \; \sgn(s_i-\sbar)+ }{(w+|s_i-\sbar|)^3} \right. \nonumber \\
 &\left.  - \frac{1}{N} \frac{ \alpha (w +|s_i -\sbar|) }{(w+|s_i-\sbar|)^3} \right]\;.
\end{align}
It follows that these off-diagonal elements approach zero as $N \rightarrow \infty$. 

The diagonal elements of the Jacobian matrix take the form
\begin{align}
J_{ii} &=  \frac{1}{(w+|s_i-\sbar|)^3} \left\{ \left[\frac{N-1}{N}\right]^2 \left[s_i-\sbar -2 \; \sgn(s_i-\sbar)\right.\right. \nonumber \\
&\left.(w+|s_i-\sbar|) 
 +  \frac{\alpha(1+\alpha)}{s_i^2} (w+|s_i-\sbar|)^2 (s_i-\sbar + |s_i-\sbar|) \right. \nonumber \\
 &\left. - w\left[\frac{N-1}{N}\right]  \left[ \frac{w+|s_i-\sbar|}{s_i} \right] \right\}\;.
\end{align}
One can show that these are all negative when $s_i = \sstar$ or $s_i \to 0^+$ and $\sbar = (1-x)\sstar$ as $N\rightarrow \infty$, given that Eq.~\ref{eq:bimod_cond} is satisfied.

Since off-diagonal elements become infinitesimal, the eigenvalues of the Jacobian matrix are determined by the diagonal elements as $N \to \infty$, and thus all eigenvalues are negative, implying linear stability of the anisogamous equilibrium.

\subsection{Stable size distributions}

\begin{figure}[th!]
	\centering
	\begin{tikzpicture}
	\node at (0,0) {\includegraphics[width=88mm]{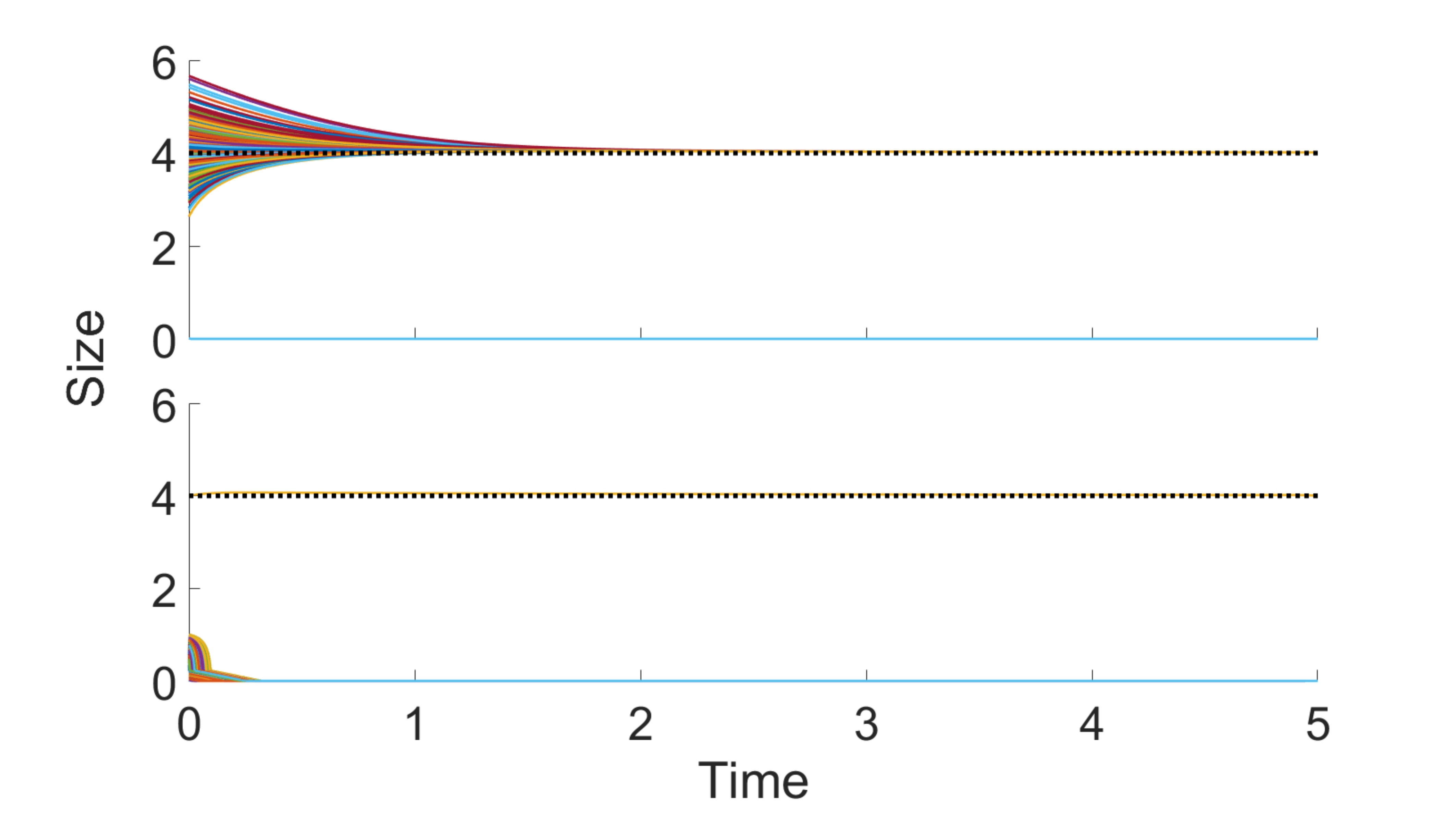}};
	\node at (2.9,1.95) {(a)};
	\node at (2.9,-0.2) {(b)};
	\end{tikzpicture}
	\caption{\textbf{Numerical test of the stability of anisogamy.} We test the stability of the anisogamous state. Panel (a) shows the large gamete group being perturbed and then returning to its equilibrium value, $s^*$. Panel (b) shows the small gamete group being perturbed and then returning to its equilibrium value zero. Dotted lines show large-gamete equilibria from theory. Panels (a) and (b) demonstrate the stability of the anisogamous equilibrium and are consistent with the asymptotic theory from Eqns.~\eqref{eq:large_stab} and \eqref{eq:small_stab}. In both panels, we set $\alpha = 1/3$, $N = 1000$, and $w=2$, with an initial fractionation $x=1/2$. }
	\label{fig:pert_tests}
\end{figure}

We perform two numerical experiments to test the stability of the anisogamous state. First, we perturb the large group from its equilibrium value $s^*$ given in \EqMid \eqref{eq:egg_size} by the amounts $\xi_i$, $i=1,\ldots,(1-x)N$, drawn from $\mathcal{N}(0,(1/2)^2)$. \FigBeg \ref{fig:pert_tests}a displays the result of the perturbation. Gamete sizes that were perturbed return their equilibrium value. Second, we perturb the small group by amounts $\xi_i$, $i=1,\ldots,xN$ drawn from $\mathcal{U}(0,1)$\footnote{We choose to perturb by the uniform distribution in order to avoid negative values.}. Similar to the first test, \FigBeg \ref{fig:pert_tests}b demonstrates that the perturbed group returns to its equilibrium value. We set $\alpha = 1/3$, $N=1000$, $w=2$, and $x=1/2$ in both simulations.

\section{Nonidentical individuals}

\begin{figure}[th!]
	\centering
	\includegraphics[width=88mm]{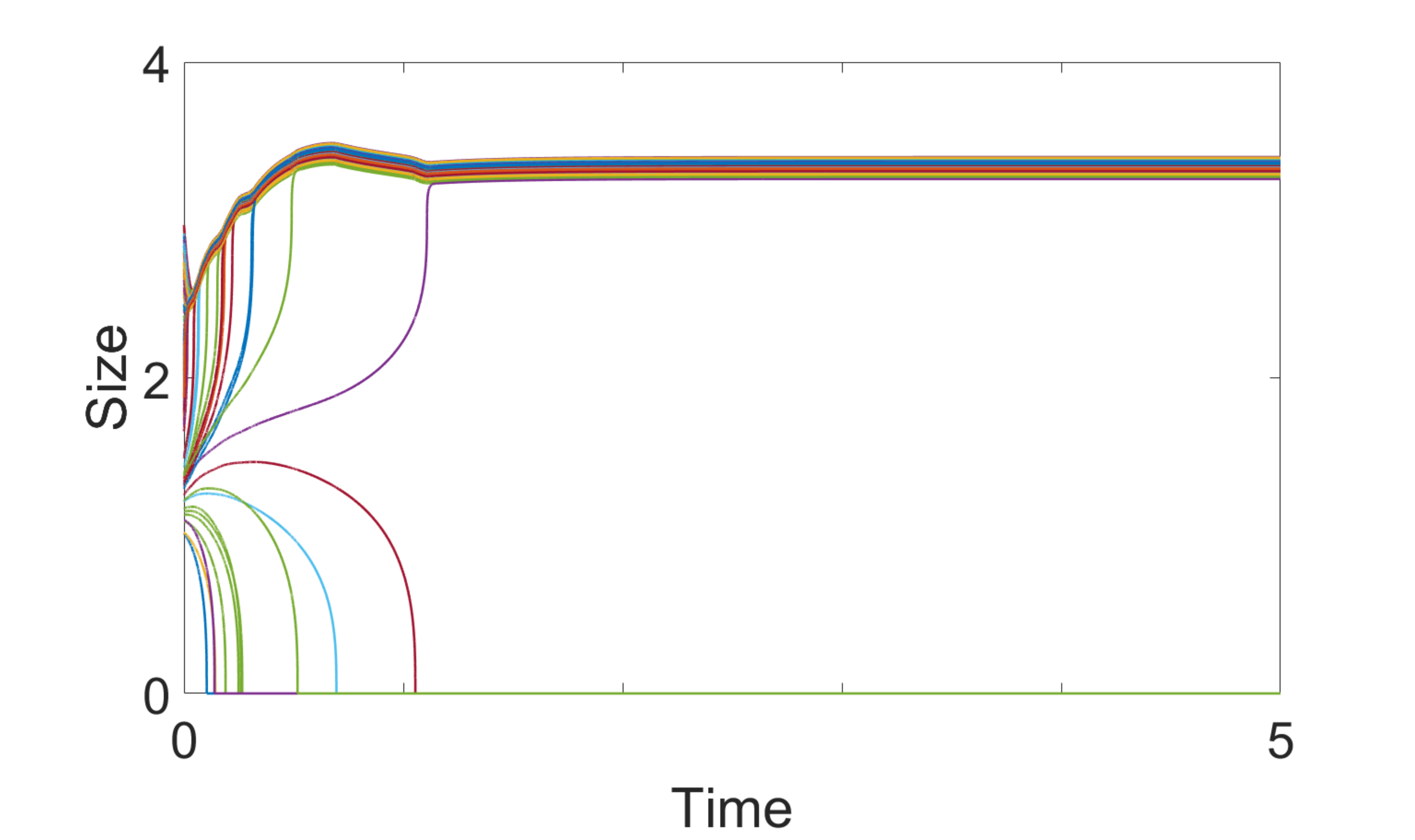}
	\caption{\textbf{Simulation with a heterogeneous population.} We display the evolution from isogamy to anisogamy for a heterogeneous population. The large gamete group widens out when adding noise to the width of the sigmoid in \EqMid \eqref{eq:gam_fit}. The final fraction of organisms that produce small gametes is $x=0.1$. For this simulation, we set $\alpha = 1$, $N = 100$, and $w$ was sampled from the distribution $\mathcal{N}(1/10,1/50^2)$.}
	\label{fig:hetero_width}
\end{figure}

Our results appear to be robust to the inclusion of natural variation among the simulated individuals. In various numerical experiments, we introduced variation in the width of the sigmoidal gamete reproductive potential function $\phig$ (see \EqMid \eqref{eq:gam_fit}), as well as in its mean, minimum, and maximum values. We also varied the multiplicative factor in the gamete production function (see \EqMid \eqref{eq:num_gam}).  In all cases, the equilibrium gamete size distribution remained qualitatively the same as in the case with identical individuals: the only change was the appearance of some variation around the expected delta function peaks (primarily the peak at $\sstar$) at equilibrium.  See Figure \ref{fig:hetero_width}.

\section{Nonzero size for small gamete group} \label{sec:nonzero_gamete}

\begin{figure}[th!]
	\centering
	\includegraphics[width=88mm]{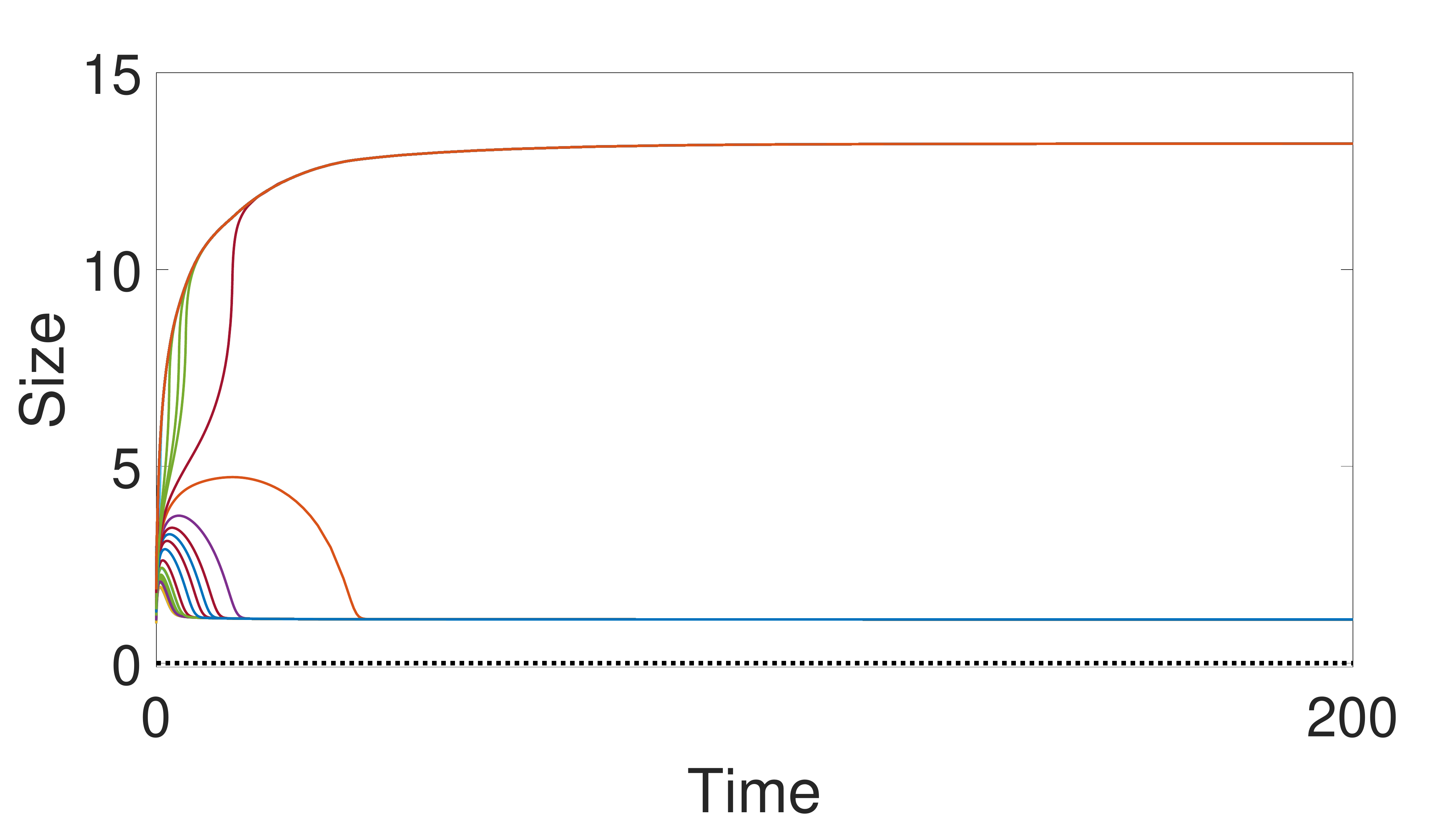}
	\caption{\textbf{Simulation with a nonzero minimum gamete size.} We display a simulation where the individual reproductive potential is multiplied by $e^{-k/s}$, where $k>0$. The initial isogamous population moves to an anisogamous population $\rho(s) = x \delta(s - \sstar_{\textrm{small}}) + (1-x) \delta(s-\sstar)$, $0<\sstar_{\textrm{small}} < \sstar$. Here, $N=100$, $w=1$, $\alpha=1$, $k=1$, and the final fraction of small gametes is $x = 0.16$. The initial isogamous population was drawn from $\mathcal{U}(1,3)$. }  
	\label{fig:varying_min_gam_size}
\end{figure}

Because reproduction requires the transfer of some minimal amount of physical material, the number of gametes cannot realistically diverge as $s \to 0^+$.  Our results, however, appear to be robust to the inclusion of a minimal viable gamete size.  In simulation, we incorporated a minimal size by multiplying the individual reproductive potential by $e^{-k/s}$, where $k>0$. This eliminated the singularity at zero and generated a point $0<\sstar_{\textrm{small}} < \sstar$ such that reproductive potential is maximized. In such simulations, the resulting equilibrium distribution was $\rho(s) = x \delta(s - \sstar_{\textrm{small}}) + (1-x) \delta(s-\sstar)$, as expected.  See Figure \ref{fig:varying_min_gam_size}.

\section{Absolute gamete fitness} \label{sec:abs_gam_fit}

\begin{figure}[th!]
	\centering
	\includegraphics[width=88mm]{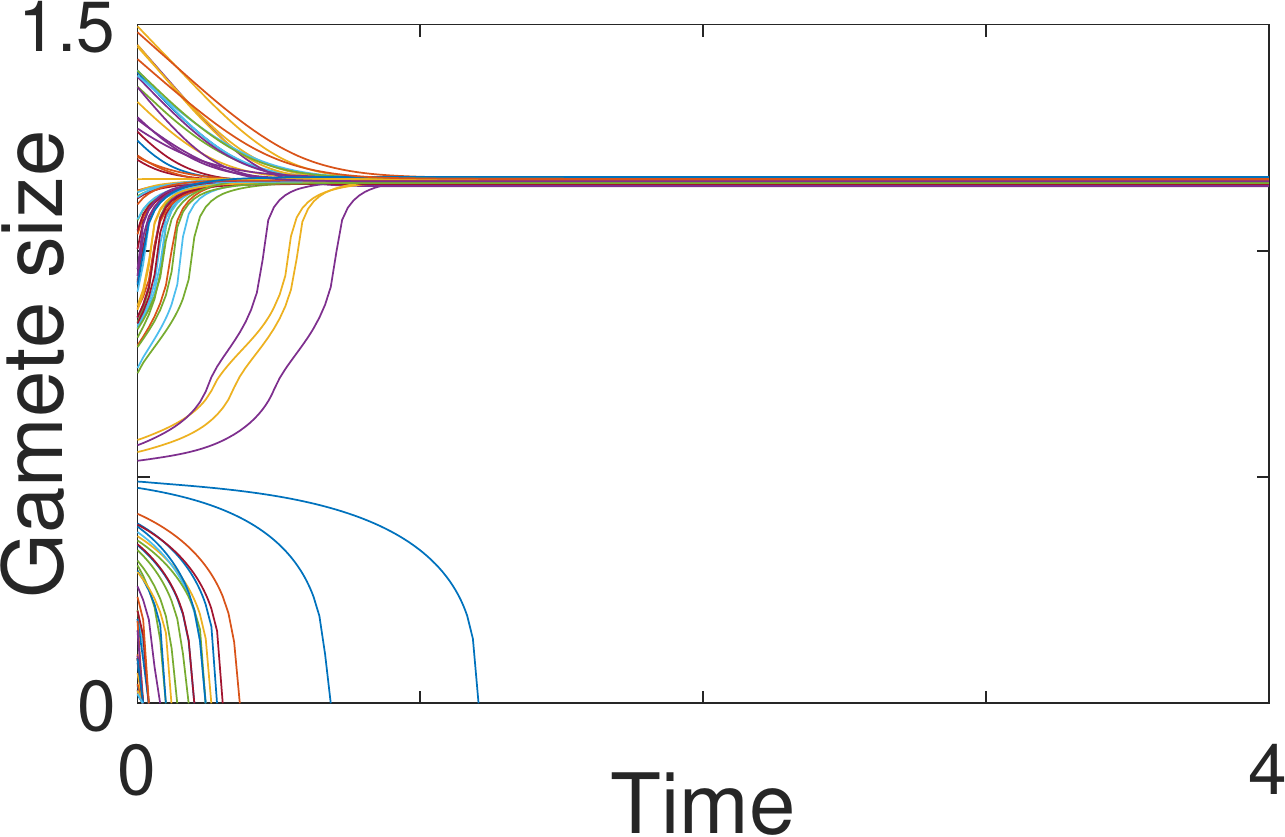}
	\caption{\textbf{Simulation with both absolute and relative reproductive potential.} In this simulation the individual reproductive potential was a weighted sum of two sigmoidal functions, one as in the \EqMid \eqref{eq:gam_fit} (i.e., centered at $\sbar$), and the other identical but centered at $c=1$.  Weight was 90\% absolute, 10\% relative. The population converges to an anisogamous state with 40\% small gametes. Here, $N=100$, $w=0.1 + \mathcal{N}(0,0.01^2)$, $\alpha=1$, and the initial population was drawn from $\mathcal{U}(0,1.5)$.}
	\label{fig:varying_abs_rel}
\end{figure}

In our model we assume that the reproductive potential of a gamete depends on its size \textit{relative} to others in the population.  In reality, there are likely some \textit{absolute} size effects that also play a role. In Figure \ref{fig:varying_abs_rel}, we numerically simulate our model with the inclusion of both absolute and relative gamete potential terms, with the results appearing to remain qualitatively unchanged. This shows that a wider range of fractionations is now possible at equilibrium; here the final fraction of small gametes $x = 0.4 > (3-\sqrt{2})/\alpha \approx 0.17$---the threshold given by \EqMid \eqref{eq:bimod_cond}.  

\section{Reproductive potential and fitness} \label{sec:potfit}

Here we briefly summarize the connection between phenotype flux and fitness that was laid out by \cite{clifton2016}---the reader is referred to that reference for greater detail. 

Many problems in evolutionary dynamics can be modeled by the replicator equation (see \cite{taylor1978evolutionary,schuster1983replicator}). In the continuum limit, it takes the form
\begin{equation}
    \frac{\partial \rho}{\partial t} =  \rho(s,t) [f(s;\rho) - \overline{f} (\rho)]\;,
    \label{eq:repeq}
\end{equation}
where $\rho$ is the probability distribution of continuous trait $s$ at time $t$, $f$ is the fitness of an organism with trait $s$ given the trait distribution, and $\overline{f} = \int_{-\infty}^{\infty} f(s;\rho) \rho(s,t)$ ds.
The trait distribution $\rho$ must always integrate to one and hence follows the continuity equation (see \cite{pedlosky1987geophysical}) 
\begin{equation}
     \frac{\partial \rho}{\partial t} =  -\nabla \cdot (\rho \frac{ds}{dt})\;, 
     \label{eq:conteq}
\end{equation}
where $\nabla \equiv \partial / \partial s$ here.

The main difference between the replicator equation and the continuity equation is that temporal changes in the trait distribution are expressed in terms of an excess fitness $f(s;\rho) - \overline{f}(\rho)$ in the former, but a phenotype flux $ds/dt$ in the latter. In our model, we take the phenotype flux as derivable from the individual reproductive potential $\phii$ as expressed in Eq.~\eqref{eq:diff_ind_fit2}.

Setting the right-hand sides of Eq.~\eqref{eq:repeq} and Eq.~\eqref{eq:conteq} equal gives 
%
\begin{equation}
    - \rho [f(s;\rho) - \overline{f} (\rho)] = \nabla \cdot (\rho \frac{ds}{dt}) =\frac{\partial \rho}{\partial s} \frac{\partial \phii}{\partial s} + \rho \frac{\partial^2 \phii}{\partial s^2}
\end{equation}
(using $\tau=1$ for convenience).  This expresses one relationship between fitness and reproductive potential. The relationship can also be expressed through an integro-differential equation for $\phii$ in terms of fitness:
\begin{equation}
    \frac{\partial \phii}{\partial s} = -\frac{1}{\rho} \int_{-\infty}^{\infty} \rho(s,t) [f(s;\rho) - \overline{f} (\rho)] ds \;.
\end{equation}






\bibliographystyle{model4-names}
\biboptions{authoryear}




\bibliography{main_jtb}

\begin{thebibliography}{38}
\expandafter\ifx\csname natexlab\endcsname\relax\def\natexlab#1{#1}\fi
\providecommand{\url}[1]{\texttt{#1}}
\providecommand{\href}[2]{#2}
\providecommand{\path}[1]{#1}
\providecommand{\DOIprefix}{doi:}
\providecommand{\ArXivprefix}{arXiv:}
\providecommand{\URLprefix}{URL: }
\providecommand{\Pubmedprefix}{pmid:}
\providecommand{\doi}[1]{\href{http://dx.doi.org/#1}{\path{#1}}}
\providecommand{\Pubmed}[1]{\href{pmid:#1}{\path{#1}}}
\providecommand{\bibinfo}[2]{#2}
\ifx\xfnm\undefined \def\xfnm[#1]{\unskip,\space#1}\fi
\bibitem[{Alberts et~al.(2002)Alberts, Johnson, Lewis, Raff, Roberts and
  Walter}]{alberts2002eggs}
\bibinfo{author}{Alberts\xfnm[ B.]}, \bibinfo{author}{Johnson\xfnm[ A.]},
  \bibinfo{author}{Lewis\xfnm[ J.]}, \bibinfo{author}{Raff\xfnm[ M.]},
  \bibinfo{author}{Roberts\xfnm[ K.]}, \bibinfo{author}{Walter\xfnm[ P.]}.
\newblock \bibinfo{title}{Eggs}.
\newblock In: \bibinfo{booktitle}{Molecular Biology of the Cell, 4th edition}.
  \bibinfo{publisher}{Garland Science}; \bibinfo{year}{2002}. .
\bibitem[{Bateman(1948)}]{bateman1948intra}
\bibinfo{author}{Bateman\xfnm[ A.J.]}.
\newblock \bibinfo{title}{Intra-sexual selection in drosophila}.
\newblock \bibinfo{journal}{Heredity}
  \bibinfo{year}{1948};\bibinfo{volume}{2}(\bibinfo{number}{3}):\bibinfo{pages}{349--368}.
\newblock \bibinfo{note}{{h}ttps://doi.org/10.1038/hdy.1948.21}.
\bibitem[{Bateman and DiMichele(1994)}]{bateman1994heterospory}
\bibinfo{author}{Bateman\xfnm[ R.M.]}, \bibinfo{author}{DiMichele\xfnm[ W.A.]}.
\newblock \bibinfo{title}{Heterospory: the most iterative key innovation in the
  evolutionary history of the plant kingdom}.
\newblock \bibinfo{journal}{Biological Reviews}
  \bibinfo{year}{1994};\bibinfo{volume}{69}(\bibinfo{number}{3}):\bibinfo{pages}{345--417}.
\newblock \bibinfo{note}{{h}ttps://doi.org/10.1111/j.1469-185X.1994.tb01276.x}.
\bibitem[{Bell(1978)}]{bell1978evolution}
\bibinfo{author}{Bell\xfnm[ G.]}.
\newblock \bibinfo{title}{The evolution of anisogamy}.
\newblock \bibinfo{journal}{Journal of Theoretical Biology}
  \bibinfo{year}{1978};\bibinfo{volume}{73}(\bibinfo{number}{2}):\bibinfo{pages}{247--270}.
\newblock \bibinfo{note}{{h}ttps://doi.org/10.1016/0022-5193(78)90189-3}.
\bibitem[{Bellastella et~al.(2010)Bellastella, Cooper, Battaglia, Str{\"o}se,
  Torres, Hellenkemper, Soler and Sinisi}]{bellastella2010dimensions}
\bibinfo{author}{Bellastella\xfnm[ G.]}, \bibinfo{author}{Cooper\xfnm[ T.G.]},
  \bibinfo{author}{Battaglia\xfnm[ M.]}, \bibinfo{author}{Str{\"o}se\xfnm[
  A.]}, \bibinfo{author}{Torres\xfnm[ I.]}, \bibinfo{author}{Hellenkemper\xfnm[
  B.]}, \bibinfo{author}{Soler\xfnm[ C.]}, \bibinfo{author}{Sinisi\xfnm[
  A.A.]}.
\newblock \bibinfo{title}{Dimensions of human ejaculated spermatozoa in
  papanicolaou-stained seminal and swim-up smears obtained from the integrated
  semen analysis system (isas{\textregistered})}.
\newblock \bibinfo{journal}{Asian Journal of Andrology}
  \bibinfo{year}{2010};\bibinfo{volume}{12}(\bibinfo{number}{6}):\bibinfo{pages}{871}.
\newblock \bibinfo{note}{Https://dx.doi.org/10.1038/aja.2010.90}.
\bibitem[{Billiard et~al.(2011)Billiard, L{\'o}pez-Villavicencio, Devier, Hood,
  Fairhead and Giraud}]{billiard2011having}
\bibinfo{author}{Billiard\xfnm[ S.]},
  \bibinfo{author}{L{\'o}pez-Villavicencio\xfnm[ M.]},
  \bibinfo{author}{Devier\xfnm[ B.]}, \bibinfo{author}{Hood\xfnm[ M.E.]},
  \bibinfo{author}{Fairhead\xfnm[ C.]}, \bibinfo{author}{Giraud\xfnm[ T.]}.
\newblock \bibinfo{title}{Having sex, yes, but with whom? inferences from fungi
  on the evolution of anisogamy and mating types}.
\newblock \bibinfo{journal}{Biological Reviews}
  \bibinfo{year}{2011};\bibinfo{volume}{86}(\bibinfo{number}{2}):\bibinfo{pages}{421--442}.
\newblock \bibinfo{note}{{h}ttps://doi.org/10.1111/j.1469-185X.2010.00153.x}.
\bibitem[{Blomqvist et~al.(1997)Blomqvist, Johansson and
  G{\"o}tmark}]{blomqvist1997parental}
\bibinfo{author}{Blomqvist\xfnm[ D.]}, \bibinfo{author}{Johansson\xfnm[ O.C.]},
  \bibinfo{author}{G{\"o}tmark\xfnm[ F.]}.
\newblock \bibinfo{title}{Parental quality and egg size affect chick survival
  in a precocial bird, the lapwing vanellus vanellus}.
\newblock \bibinfo{journal}{Oecologia}
  \bibinfo{year}{1997};\bibinfo{volume}{110}(\bibinfo{number}{1}):\bibinfo{pages}{18--24}.
\newblock \bibinfo{note}{{h}ttps://doi.org/10.1007/s004420050128}.
\bibitem[{Blute(2013)}]{blute2013evolution}
\bibinfo{author}{Blute\xfnm[ M.]}.
\newblock \bibinfo{title}{The evolution of anisogamy: more questions than
  answers}.
\newblock \bibinfo{journal}{Biological Theory}
  \bibinfo{year}{2013};\bibinfo{volume}{7}(\bibinfo{number}{1}):\bibinfo{pages}{3--9}.
\newblock \bibinfo{note}{Https://doi.org/10.1007/s13752-012-0060-4}.
\bibitem[{Bonsall(2006)}]{bonsall2006evolution}
\bibinfo{author}{Bonsall\xfnm[ M.B.]}.
\newblock \bibinfo{title}{The evolution of anisogamy: The adaptive significance
  of damage, repair and mortality}.
\newblock \bibinfo{journal}{Journal of Theoretical Biology}
  \bibinfo{year}{2006};\bibinfo{volume}{238}(\bibinfo{number}{1}):\bibinfo{pages}{198--210}.
\newblock \bibinfo{note}{Https://doi.org/10.1016/j.jtbi.2005.05.007}.
\bibitem[{Brown et~al.(2002)Brown, Gupta, Li, Milne, Restrepo and
  West}]{brown2002fractal}
\bibinfo{author}{Brown\xfnm[ J.H.]}, \bibinfo{author}{Gupta\xfnm[ V.K.]},
  \bibinfo{author}{Li\xfnm[ B.L.]}, \bibinfo{author}{Milne\xfnm[ B.T.]},
  \bibinfo{author}{Restrepo\xfnm[ C.]}, \bibinfo{author}{West\xfnm[ G.B.]}.
\newblock \bibinfo{title}{The fractal nature of nature: power laws, ecological
  complexity and biodiversity}.
\newblock \bibinfo{journal}{Philosophical Transactions of the Royal Society of
  London Series B: Biological Sciences}
  \bibinfo{year}{2002};\bibinfo{volume}{357}(\bibinfo{number}{1421}):\bibinfo{pages}{619--626}.
\newblock \bibinfo{note}{{h}ttps://doi.org/10.1098/rstb.2001.0993}.
\bibitem[{Bulmer and Parker(2002)}]{bulmer2002evolution}
\bibinfo{author}{Bulmer\xfnm[ M.]}, \bibinfo{author}{Parker\xfnm[ G.A.]}.
\newblock \bibinfo{title}{The evolution of anisogamy: a game-theoretic
  approach}.
\newblock \bibinfo{journal}{Proceedings of the Royal Society of London Series
  B: Biological Sciences}
  \bibinfo{year}{2002};\bibinfo{volume}{269}(\bibinfo{number}{1507}):\bibinfo{pages}{2381--2388}.
\newblock \bibinfo{note}{{h}ttps://doi.org/10.1098/rspb.2002.2161}.
\bibitem[{Charlesworth(1978)}]{charlesworth1978population}
\bibinfo{author}{Charlesworth\xfnm[ B.]}.
\newblock \bibinfo{title}{The population genetics of anisogamy}.
\newblock \bibinfo{journal}{Journal of theoretical biology}
  \bibinfo{year}{1978};\bibinfo{volume}{73}(\bibinfo{number}{2}):\bibinfo{pages}{347--357}.
\bibitem[{Clauset et~al.(2009)Clauset, Shalizi and Newman}]{clauset2009power}
\bibinfo{author}{Clauset\xfnm[ A.]}, \bibinfo{author}{Shalizi\xfnm[ C.R.]},
  \bibinfo{author}{Newman\xfnm[ M.E.]}.
\newblock \bibinfo{title}{Power-law distributions in empirical data}.
\newblock \bibinfo{journal}{SIAM Review}
  \bibinfo{year}{2009};\bibinfo{volume}{51}(\bibinfo{number}{4}):\bibinfo{pages}{661--703}.
\newblock \bibinfo{note}{{h}ttps://doi.org/10.1137/070710111}.
\bibitem[{Clifton et~al.(2016)Clifton, Braun and Abrams}]{clifton2016}
\bibinfo{author}{Clifton\xfnm[ S.M.]}, \bibinfo{author}{Braun\xfnm[ R.I.]},
  \bibinfo{author}{Abrams\xfnm[ D.M.]}.
\newblock \bibinfo{title}{Handicap principle implies emergence of dimorphic
  ornaments}.
\newblock \bibinfo{journal}{Proceedings of the Royal Society B: Biological
  Sciences}
  \bibinfo{year}{2016};\bibinfo{volume}{283}(\bibinfo{number}{1843}):\bibinfo{pages}{20161970}.
\newblock \bibinfo{note}{{h}ttps://doi.org/10.1098/rspb.2016.1970}.
\bibitem[{Cox and Sethian(1985)}]{cox1985gamete}
\bibinfo{author}{Cox\xfnm[ P.A.]}, \bibinfo{author}{Sethian\xfnm[ J.A.]}.
\newblock \bibinfo{title}{Gamete motion, search, and the evolution of
  anisogamy, oogamy, and chemotaxis}.
\newblock \bibinfo{journal}{The American Naturalist}
  \bibinfo{year}{1985};\bibinfo{volume}{125}(\bibinfo{number}{1}):\bibinfo{pages}{74--101}.
\newblock \bibinfo{note}{{h}ttps://doi.org/10.1086/284329}.
\bibitem[{Erikstad et~al.(1998)Erikstad, Tveraa and
  Bustnes}]{erikstad1998significance}
\bibinfo{author}{Erikstad\xfnm[ K.E.]}, \bibinfo{author}{Tveraa\xfnm[ T.]},
  \bibinfo{author}{Bustnes\xfnm[ J.O.]}.
\newblock \bibinfo{title}{Significance of intraclutch egg-size variation in
  common eider: the role of egg size and quality of ducklings}.
\newblock \bibinfo{journal}{Journal of Avian Biology}
  \bibinfo{year}{1998};:\bibinfo{pages}{3--9}\bibinfo{note}{{h}ttps://doi.org/10.2307/3677334}.
\bibitem[{Haig and Westoby(1988)}]{haig1988model}
\bibinfo{author}{Haig\xfnm[ D.]}, \bibinfo{author}{Westoby\xfnm[ M.]}.
\newblock \bibinfo{title}{A model for the origin of heterospory}.
\newblock \bibinfo{journal}{Journal of Theoretical Biology}
  \bibinfo{year}{1988};\bibinfo{volume}{134}(\bibinfo{number}{2}):\bibinfo{pages}{257--272}.
\newblock \bibinfo{note}{{h}ttps://doi.org/10.1016/S0022-5193(88)80203-0}.
\bibitem[{Hamilton(1967)}]{hamilton1967extraordinary}
\bibinfo{author}{Hamilton\xfnm[ W.D.]}.
\newblock \bibinfo{title}{Extraordinary sex ratios}.
\newblock \bibinfo{journal}{Science}
  \bibinfo{year}{1967};\bibinfo{volume}{156}(\bibinfo{number}{3774}):\bibinfo{pages}{477--488}.
\newblock \bibinfo{note}{{h}ttps://www.doi.org/10.1126/science.156.3774.477}.
\bibitem[{Hayward and Gillooly(2011)}]{hayward2011cost}
\bibinfo{author}{Hayward\xfnm[ A.]}, \bibinfo{author}{Gillooly\xfnm[ J.F.]}.
\newblock \bibinfo{title}{The cost of sex: quantifying energetic investment in
  gamete production by males and females}.
\newblock \bibinfo{journal}{PLoS One}
  \bibinfo{year}{2011};\bibinfo{volume}{6}(\bibinfo{number}{1}):\bibinfo{pages}{e16557}.
\newblock \bibinfo{note}{{h}ttps://doi.org/10.1371/journal.pone.0016557}.
\bibitem[{Hedges(2008)}]{hedges2008lower}
\bibinfo{author}{Hedges\xfnm[ S.B.]}.
\newblock \bibinfo{title}{At the lower size limit in snakes: two new species of
  threadsnakes (squamata: Leptotyphlopidae: Leptotyphlops) from the lesser
  antilles}.
\newblock \bibinfo{journal}{Zootaxa}
  \bibinfo{year}{2008};\bibinfo{volume}{1841}(\bibinfo{number}{1}):\bibinfo{pages}{1--30}.
\bibitem[{Hurst(1990)}]{hurst1990parasite}
\bibinfo{author}{Hurst\xfnm[ L.D.]}.
\newblock \bibinfo{title}{Parasite diversity and the evolution of diploidy,
  multicellularity and anisogamy}.
\newblock \bibinfo{journal}{Journal of Theoretical Biology}
  \bibinfo{year}{1990};\bibinfo{volume}{144}(\bibinfo{number}{4}):\bibinfo{pages}{429--443}.
\newblock \bibinfo{note}{{h}ttps://doi.org/10.1016/S0022-5193(05)80085-2}.
\bibitem[{Johnson et~al.(1983)Johnson, Petty and Neaves}]{johnson1983further}
\bibinfo{author}{Johnson\xfnm[ L.]}, \bibinfo{author}{Petty\xfnm[ C.S.]},
  \bibinfo{author}{Neaves\xfnm[ W.B.]}.
\newblock \bibinfo{title}{Further quantification of human spermatogenesis: germ
  cell loss during postprophase of meiosis and its relationship to daily sperm
  production}.
\newblock \bibinfo{journal}{Biology of Reproduction}
  \bibinfo{year}{1983};\bibinfo{volume}{29}(\bibinfo{number}{1}):\bibinfo{pages}{207--215}.
\newblock \bibinfo{note}{Https://doi.org/10.1095/biolreprod29.1.207}.
\bibitem[{Kalmus(1932)}]{kalmus1932erhaltungswert}
\bibinfo{author}{Kalmus\xfnm[ H.]}.
\newblock \bibinfo{title}{{\"U}ber den erhaltungswert der ph{\"a}notypischen
  (morphologischen) anisogamie und die entstehung der ersten
  geschlechtsunterschiede}.
\newblock \bibinfo{journal}{Biologisches Zentralblatt}
  \bibinfo{year}{1932};\bibinfo{volume}{52}:\bibinfo{pages}{716--736}.
\bibitem[{Kalmus and Smith(1960)}]{kalmus1960evolutionary}
\bibinfo{author}{Kalmus\xfnm[ H.]}, \bibinfo{author}{Smith\xfnm[ C.]}.
\newblock \bibinfo{title}{Evolutionary origin of sexual differentiation and the
  sex-ratio}.
\newblock \bibinfo{journal}{Nature}
  \bibinfo{year}{1960};\bibinfo{volume}{186}(\bibinfo{number}{4730}):\bibinfo{pages}{1004--1006}.
\newblock \bibinfo{note}{{h}ttps://doi.org/10.1038/1861004a0}.
\bibitem[{Krist(2011)}]{krist2011egg}
\bibinfo{author}{Krist\xfnm[ M.]}.
\newblock \bibinfo{title}{Egg size and offspring quality: a meta-analysis in
  birds}.
\newblock \bibinfo{journal}{Biological Reviews}
  \bibinfo{year}{2011};\bibinfo{volume}{86}(\bibinfo{number}{3}):\bibinfo{pages}{692--716}.
\newblock \bibinfo{note}{Https://doi.org/10.1111/j.1469-185X.2010.00166.x}.
\bibitem[{Kuees(2015)}]{kuees2015two}
\bibinfo{author}{Kuees\xfnm[ U.]}.
\newblock \bibinfo{title}{From two to many: multiple mating types in
  basidiomycetes}.
\newblock \bibinfo{journal}{Fungal Biology Reviews}
  \bibinfo{year}{2015};\bibinfo{volume}{29}(\bibinfo{number}{3-4}):\bibinfo{pages}{126--166}.
\newblock \bibinfo{note}{{h}ttps://doi.org/10.1016/j.fbr.2015.11.001}.
\bibitem[{Kues and Casselton(1993)}]{kues1993origin}
\bibinfo{author}{Kues\xfnm[ U.]}, \bibinfo{author}{Casselton\xfnm[ L.A.]}.
\newblock \bibinfo{title}{The origin of multiple mating types in mushrooms}.
\newblock \bibinfo{journal}{Journal of Cell Science}
  \bibinfo{year}{1993};\bibinfo{volume}{104}(\bibinfo{number}{2}):\bibinfo{pages}{227--230}.
\bibitem[{Lehtonen et~al.(2016)Lehtonen, Parker and
  Sch{\"a}rer}]{lehtonen2016anisogamy}
\bibinfo{author}{Lehtonen\xfnm[ J.]}, \bibinfo{author}{Parker\xfnm[ G.A.]},
  \bibinfo{author}{Sch{\"a}rer\xfnm[ L.]}.
\newblock \bibinfo{title}{Why anisogamy drives ancestral sex roles}.
\newblock \bibinfo{journal}{Evolution}
  \bibinfo{year}{2016};\bibinfo{volume}{70}(\bibinfo{number}{5}):\bibinfo{pages}{1129--1135}.
\newblock \bibinfo{note}{{h}ttps://doi.org/10.1111/evo.12926}.
\bibitem[{Nager et~al.(2000)Nager, Monaghan and Houston}]{nager2000within}
\bibinfo{author}{Nager\xfnm[ R.G.]}, \bibinfo{author}{Monaghan\xfnm[ P.]},
  \bibinfo{author}{Houston\xfnm[ D.C.]}.
\newblock \bibinfo{title}{Within-clutch trade-offs between the number and
  quality of eggs: experimental manipulations in gulls}.
\newblock \bibinfo{journal}{Ecology}
  \bibinfo{year}{2000};\bibinfo{volume}{81}(\bibinfo{number}{5}):\bibinfo{pages}{1339--1350}.
\bibitem[{Newman(2005)}]{newman2005power}
\bibinfo{author}{Newman\xfnm[ M.E.]}.
\newblock \bibinfo{title}{Power laws, pareto distributions and zipf's law}.
\newblock \bibinfo{journal}{Contemporary Physics}
  \bibinfo{year}{2005};\bibinfo{volume}{46}(\bibinfo{number}{5}):\bibinfo{pages}{323--351}.
\newblock \bibinfo{note}{{h}ttps://doi.org/10.1080/00107510500052444}.
\bibitem[{Parker et~al.(1972)Parker, Baker and Smith}]{parker1972origin}
\bibinfo{author}{Parker\xfnm[ G.A.]}, \bibinfo{author}{Baker\xfnm[ R.R.]},
  \bibinfo{author}{Smith\xfnm[ V.]}.
\newblock \bibinfo{title}{The origin and evolution of gamete dimorphism and the
  male-female phenomenon}.
\newblock \bibinfo{journal}{Journal of Theoretical Biology}
  \bibinfo{year}{1972};\bibinfo{volume}{36}(\bibinfo{number}{3}):\bibinfo{pages}{529--553}.
\newblock \bibinfo{note}{{h}ttps://doi.org/10.1016/0022-5193(72)90007-0}.
\bibitem[{Pedlosky(1987)}]{pedlosky1987geophysical}
\bibinfo{author}{Pedlosky\xfnm[ J.]}.
\newblock \bibinfo{title}{Geophysical fluid dynamics};
  \bibinfo{publisher}{Springer}.
\newblock \bibinfo{edition}{2nd} ed.; p.~\bibinfo{pages}{10}.
\newblock \bibinfo{note}{{h}ttps://doi.org/10.1007/978-1-4612-4650-3}.
\bibitem[{Schuster and Sigmund(1983)}]{schuster1983replicator}
\bibinfo{author}{Schuster\xfnm[ P.]}, \bibinfo{author}{Sigmund\xfnm[ K.]}.
\newblock \bibinfo{title}{Replicator dynamics}.
\newblock \bibinfo{journal}{Journal of Theoretical Biology}
  \bibinfo{year}{1983};\bibinfo{volume}{100}(\bibinfo{number}{3}):\bibinfo{pages}{533--538}.
\newblock \bibinfo{note}{{h}ttps://doi.org/10.1016/0022-5193(83)90445-9}.
\bibitem[{Scudo(1967)}]{scudo1967adaptive}
\bibinfo{author}{Scudo\xfnm[ F.M.]}.
\newblock \bibinfo{title}{The adaptive value of sexual dimorphism: I,
  anisogamy}.
\newblock \bibinfo{journal}{Evolution}
  \bibinfo{year}{1967};:\bibinfo{pages}{285--291}\bibinfo{note}{{h}ttps://doi.org/10.1111/j.1558-5646.1967.tb00156.x}.
\bibitem[{da~Silva(2018)}]{da2018evolution}
\bibinfo{author}{da~Silva\xfnm[ J.]}.
\newblock \bibinfo{title}{The evolution of sexes: A specific test of the
  disruptive selection theory}.
\newblock \bibinfo{journal}{Ecology and evolution}
  \bibinfo{year}{2018};\bibinfo{volume}{8}(\bibinfo{number}{1}):\bibinfo{pages}{207--219}.
\bibitem[{Taylor and Jonker(1978)}]{taylor1978evolutionary}
\bibinfo{author}{Taylor\xfnm[ P.D.]}, \bibinfo{author}{Jonker\xfnm[ L.B.]}.
\newblock \bibinfo{title}{Evolutionary stable strategies and game dynamics}.
\newblock \bibinfo{journal}{Mathematical Biosciences}
  \bibinfo{year}{1978};\bibinfo{volume}{40}(\bibinfo{number}{1-2}):\bibinfo{pages}{145--156}.
\newblock \bibinfo{note}{{h}ttps://doi.org/10.1016/0025-5564(78)90077-9}.
\bibitem[{Valkama et~al.(2002)Valkama, Korpim{\"a}ki, Wiehn and
  Pakkanen}]{valkama2002inter}
\bibinfo{author}{Valkama\xfnm[ J.]}, \bibinfo{author}{Korpim{\"a}ki\xfnm[ E.]},
  \bibinfo{author}{Wiehn\xfnm[ J.]}, \bibinfo{author}{Pakkanen\xfnm[ T.]}.
\newblock \bibinfo{title}{Inter-clutch egg size variation in kestrels falco
  tinnunculus: seasonal decline under fluctuating food conditions}.
\newblock \bibinfo{journal}{Journal of Avian Biology}
  \bibinfo{year}{2002};\bibinfo{volume}{33}(\bibinfo{number}{4}):\bibinfo{pages}{426--432}.
\newblock \bibinfo{note}{{h}ttps://doi.org/10.1034/j.1600-048X.2002.02875.x}.
\bibitem[{Wallace and Kelsey(2010)}]{wallace2010human}
\bibinfo{author}{Wallace\xfnm[ W.H.B.]}, \bibinfo{author}{Kelsey\xfnm[ T.W.]}.
\newblock \bibinfo{title}{Human ovarian reserve from conception to the
  menopause}.
\newblock \bibinfo{journal}{PloS One}
  \bibinfo{year}{2010};\bibinfo{volume}{5}(\bibinfo{number}{1}).
\newblock \bibinfo{note}{Https://doi.org/10.1371/journal.pone.0008772}.

\end{thebibliography}

\end{document}